\documentclass[12pt]{iopart}

\usepackage{graphicx}
\usepackage{units}
\usepackage{amsfonts}
\usepackage{amssymb}

\usepackage{cite}

\begin{document}

\title[Ternary surface alloys]{Tuning independently Fermi energy and
  spin splitting in Rashba systems: Ternary surface alloys on Ag(111)}

\author{H Mirhosseini, A Ernst, S Ostanin, and J Henk}
\address{Max-Planck-Institut f\"ur Mikrostrukturphysik, Weinberg
  2, D-06120 Halle (Saale), Germany}
\ead{\mailto{hossein@mpi-halle.de} (H Mirhosseini)}


\begin{abstract}
  By detailed first-principles calculations we show that the Fermi
  energy and the Rashba splitting in disordered ternary surface alloys
  Bi$_{x}$Pb$_{y}$Sb$_{1-x-y}$/Ag(111) can be independently tuned by
  choosing the concentrations $x$ and $y$ of Bi and Pb, respectively.
  The findings are explained by three fundamental mechanisms, namely
  the relaxation of the adatoms, the strength of the atomic spin-orbit
  coupling, and band filling.  By mapping the Rashba characteristics,
  i.\,e.\ the splitting $k_{\mathrm{R}}$ and the Rashba energy
  $E_{\mathrm{R}}$, and the Fermi energy of the surface states in the
  complete range of concentrations, we find that these quantities
  depend monotonically on $x$ and $y$, with a very few exceptions.
  Our results suggest to investigate experimentally effects which rely
  on the Rashba spin-orbit coupling in dependence on spin-orbit
  splitting and band filling.
\end{abstract}

\maketitle

\section{Introduction}
\label{sec:Intro}
In the emerging field of spin electronics, proposed device
applications often utilize the Rashba effect \cite{Bychkov84a} in a
two-dimensional electron gas (2DEG). A prominent example is the spin
field-effect transistor \cite{Datta90b} in which the spin-orbit (SO)
interaction in the 2DEG is controlled via a gate voltage
\cite{Nitta97, Koga02}. Other examples are a high critical superconducting temperature which shows up in materials with a sizable spin-orbit interaction \cite{Cappelluti07} and the spin Hall effect \cite{Hirsch99a, Kato04,Saraga05,Valenzuela06}.

The Rashba effect relies on breaking the inversion symmetry of the
system and, consequently, shows up in semiconductor heterostructures
and at surfaces. The breaking of the inversion symmetry results---via
the spin-orbit coupling---in a splitting in the dispersion relation of
electronic states which are confined to the interface
\cite{Bychkov84a}. In a simple model for a two-dimensional electron
gas, a potential in $z$ direction confines the electrons to the $xy$
plane. The Hamiltonian of the spin-orbit coupling can thus be written
as
\begin{equation}
  \label{eq:2DSOC}
  \hat{H}_{\mathrm{so}} =
  \gamma_{\mathrm{\mathrm{R}}} (\sigma_{x}\partial_{y} - \sigma_{y}\partial_{x}),
\end{equation}
where the strength of the SO interaction is quantified by the Rashba
parameter $\gamma_{\mathrm{\mathrm{R}}}$. Employing a plane-wave
\textit{ansatz} yields the dispersion relation
\begin{equation}
  E_{\pm}(\vec{k}_{\parallel})
  = 
  E_{0}
  +
  \frac{\hbar^{2} k_{\parallel}^{2}}{2m^{\star}}
  \pm
  \gamma_{\mathrm{\mathrm{R}}} |\vec{k}_{\parallel}|,
  \label{eq:dispersion}
\end{equation}
where $m^{\star}$ is the effective electron mass. The split electronic
states are labeled by $+$ and $-$; their spins lie within the $xy$
plane, are aligned in opposite directions, and are perpendicular to
the wave vector $\vec{k}_{\parallel}$.

In a real system, the Rashba parameter $\gamma_{\mathrm{\mathrm{R}}}$
comprises effectively two contributions \cite{Petersen00}. The
`atomic' contribution is due to the strong potential of the ions (atomic spin-orbit coupling). The
`confinement' contribution is due to the structural inversion asymmetry which can be viewed as the gradient of the confinement
potential in $z$ direction.  The larger this gradient and the atomic
spin-orbit parameter, the larger $\gamma_{\mathrm{\mathrm{R}}}$ and
the splitting
\begin{equation}
  k_{\mathrm{\mathrm{R}}}  = \frac{\left| m^{\star}
    \right|\gamma_{\mathrm{\mathrm{R}}}}{\hbar^{2}},
  \label{eq:Deltak}
\end{equation}
which is defined as the shift of the band extremum off the Brillouin
zone center ($\vec{k}_{\parallel} = 0$). Another quantification of the
splitting is the Rashba energy
\begin{equation}
  E_{\mathrm{\mathrm{R}}}  =
  -\frac{\hbar^{2} k_{\mathrm{\mathrm{R}}}^{2}}{2 m^{\star}}
  =
  -\frac{m^{\star} \gamma_{\mathrm{\mathrm{R}}}^{2}}{2 \hbar^{2}},
  \label{eq:ERashba}
\end{equation}
that is the energy of the band extremum with respect to the energy
$E_{0}$ for which the bands cross at $\vec{k}_{\parallel} = 0$.

The above dispersion relation suggests to distinguish two energy ranges. Region I is defined as the energy range between $E_{0}$ and the band
extrema ($E \in [E_{0} - E_{\mathrm{R}}, E_{0}]$ for positive
$m^{\star}$ or $E \in [E_{0}, E_{0} + E_{\mathrm{R}}]$ for negative
$m^{\star}$) \cite{Ast08}. Region II comprises the other range of band
energies ($E > E_{0}$ for positive $m^{\star}$ or $E < E_{0}$ for
negative $m^{\star}$). The density of states in region I is singular at the band extrema and decreases towards $E_{0}$ while in region II it is constant.

In view of designing device applications and investigating fundamental
effects, it is desirable to tune both the strength
$\gamma_{\mathrm{R}}$ of the Rashba spin-orbit coupling and the Fermi
energy $E_{\mathrm{F}}$ of the 2DEG\@. In a semiconductor
heterostructure, this can be achieved by an external gate voltage and
by doping of the semiconductor host materials. At a surface, these
quantities can be affected by adsorption of adatoms \cite{Forster04,
  Moreschini08a}, by surface alloying \cite{Ast08}, and by changing
the thickness of buffer layers (e.\,g.\ in Bi/(Ag)$_{n}$/Si(111)
\cite{Frantzeskakis08}). Recently, a ferroelectric control has been
proposed \cite{Mirhosseini10}.

Surface states in surface alloys show an unmatched Rashba splitting
\cite{Ast07a}, as has been investigated in detail by scanning
tunneling microscopy as well as by spin- and angle-resolved
photoelectron spectroscopy. They are convenient systems for testing
fundamental Rashba-based effects.  The ordered surface alloys
Bi/Ag(111), Pb/Ag(111), and Sb/Ag(111) have been investigated
by first-principles calculations and in experiments \cite{Ast07a, Bihlmayer07,
  Moreschini09a}.  These three systems differ with respect to their
Rashba characteristics $k_{\mathrm{R}}$ and $E_{\mathrm{R}}$, and by
$E_{0}$.  The challenge we are dealing with is how to tune these
properties \emph{independently}.

The basic idea is as follows. Bi/Ag(111) has a large splitting and occupied
$sp_{z}$ surface states, while Pb/Ag(111) has a large splitting and
unoccupied $sp_{z}$ surface states. In a disordered binary alloy
Bi$_{x}$Pb$_{1-x}$/Ag(111) the Fermi energy can be tuned by the
concentration $x$, while keeping a large spin splitting. In
contrast, Sb/Ag(111) has occupied surface states with almost the same
binding energy as those in Bi/Ag(111) but a minor splitting. This
allows to tune mainly the spin splitting but keeping the Fermi energy
in Bi$_{x}$Sb$_{1-x}$/Ag(111). Thus, by an appropriate choice of
concentrations $x$ and $y$ in a ternary alloy
Bi$_{x}$Pb$_{y}$Sb$_{1-x-y}$/Ag(111) we expect to tune the Fermi
energy and the splitting \emph{independently}. In particular, one
could access the region I between $E_{0}$ and the band maxima which is
important for high-temperature superconductivity \cite{Cappelluti07}.

We report on a first-principles investigation of disordered surface
alloys Bi$_{x}$Pb$_{y}$Sb$_{1-x-y}$/Ag(111), performed along the
successful line of our previous works on both ordered and disordered
alloys \cite{Moreschini09a, Ast07a, Ast08}.  Since all ordered and
disordered binary alloys show a $\sqrt{3} \times \sqrt{3} R30^{\circ}$
surface reconstruction, we assume this geometry also for the ternary
alloys.  The resulting substitutional disorder is described within the
coherent potential approximation.

The paper is organized as follows. Our computational approach is
sketched in \sref{sec:computational}. The results are discussed in
\sref{sec:results}, for binary alloys in \sref{sec:binary} and for
ternary alloys in \sref{sec:ternary}. We give conclusions in
\sref{sec:conclusions}.

\section{Computational aspects}
\label{sec:computational}
We rely on our successful multi-code approach, based on the local
density approximation to density functional theory. Because this is
described in detail elsewhere \cite{Mirhosseini10}, we deliberately
sketch it in this paper.

The surface relaxations of ordered surface alloys were determined
using the Vienna \textit{Ab-initio} Simulation Package (VASP)
\cite{Kresse96}, well-known for providing precise total energies and
forces. The relaxed structural parameters serve as input for first-principles
multiple-scattering calculations.  Our Korringa-Kohn-Rostoker (KKR)
method already proved successful for relativistic electronic-structure
computations of Rashba systems \cite{Henk04a, Frantzeskakis08}.

The central quantity in multiple-scattering theory is the Green
function \cite{Zabloudil05}
\begin{eqnarray}
  \label{eq:greens-function}
  G(\vec{r}_{n}, {\vec{r}_{m}}^{\prime}; E, \vec{k})
   & = &
   \sum_{\Lambda \Lambda'} Z^{n}_{\Lambda}(\vec{r}_{n}; E)
   \tau_{\Lambda \Lambda'}^{nm}(E, \vec{k})
   Z^{m}_{\Lambda'}({\vec{r}_{m}}^{\prime}; E)^{\star}
   \nonumber\\
   & - &
   \delta^{nm} \sum_{\Lambda} Z_{\Lambda}^{n}(\vec{r}_{<}; E)
   J_{\Lambda}^{n}(\vec{r}_{>}; E)^{\star},
\end{eqnarray}
where $Z$ and $J$ are regular and irregular scattering solutions of
sites $n$ and $m$ at energy $E$ and wavevector $\vec{k}$,
respectively. $\vec{r}_{n}$ is taken with respect to the position
$\vec{R}_{n}$ of site $n$ ($\vec{r}_{n} = \vec{r} - \vec{R}_{n}$).
$r_{<}$ ($r_{>}$) is the lesser (larger) of $r_{n}$ and $r_{n}'$.
$\Lambda = (\kappa, \mu)$ comprises the relativistic
spin-angular-momentum quantum numbers \cite{Zabloudil05}.  The
scattering-path operator $\tau$ is obtained in standard KKR from the
so-called KKR equation \cite{Zabloudil05}, or in layer-KKR from the
Dyson equation for the Green function \cite{Henk01c}.

The local electronic structure is analyzed in terms of the spectral
density
\begin{equation}
  N^{n}(E; \vec{k})
  =
  -\frac{1}{\pi}
  \Im \mathrm{Tr}\ G(\vec{r}_{n}, \vec{r}_{n}; E, \vec{k}).
  \label{eq:sd}
\end{equation}
By taking appropriate decompositions of the trace, the spectral
density provides information on spin polarization and orbital
composition of the electronic states.

Substitutional ternary alloys Bi$_{x}$Pb$_{y}$Sb$_{1-x-y}$/Ag(111) are
described within the coherent potential approximation (KKR-CPA), in
which short-range order is neglected. From the agreement of the
theoretical data with their experimental counterparts for the binary
alloys Bi$_{x}$Pb$_{1-x}$/Ag(111) \cite{Ast08}, we conclude that
short-range order is of minor importance in these systems. Hence, we
applied the KKR-CPA also for the ternary alloys.

The effect of the disorder can be understood as a self-energy
\cite{Weinberger90}. As a consequence, the spectral density of the
disordered alloys becomes blurred (or smeared out) as compared to that
of the ordered alloys.

\section{Results and discussion}
\label{sec:results}

\subsection{Geometry}
Relaxations have been determined by VASP for the ordered
alloys, with $\sqrt{3} \times \sqrt{3} R30^{\circ}$ reconstruction and
face-centered-cubic (fcc) stacking (VASP cannot treat substitutional
disorder within the CPA). It turns out that the relaxations of Sb, Bi,
and Pb are in accord with their atomic radii.  To be more precise, the
outward relaxations are $9.6$, $15$, and $18$ percent of the Ag(111)
bulk interlayer spacing ($\unit[2.33]{\AA}$), respectively, with
respect to the positions of the Ag atoms in the topmost layer. Being
negligibly small, in-plane displacements of Ag atoms are not
considered.

Since all ordered and disordered binary alloys show a $\sqrt{3} \times
\sqrt{3} R30^{\circ}$ surface reconstruction, we assume this geometry
also for the ternary alloys. The relaxations of the disordered surface
alloys were linearly interpolated, in dependence on the concentrations
of the constituting elements Bi, Pb, and Sb. This assumption is within
the spirit of the CPA; being a mean-field theory, a disordered system
is described by an effective medium. Likewise the relaxation should be
taken as a concentration-weighted average. We are aware, however, that
in real samples, the relaxations of the constituting individual atoms
could differ, as might be checked by scanning tunneling microscopy.

\subsection{Mechanisms which influence the Rashba-split surface states}
Before presenting details of our calculations, a brief discussion of
the general trends and mechanisms is in order. For tuning the Fermi
energy and spin splitting independently, the underlying mechanisms
should be independent as well.

A first mechanism is relaxation. The outward relaxations of Sb, Pb,
and Bi are in accord with their atomic radii; the larger the atomic
radius, the larger the outward relaxation. The relaxation is
accompanied by a charge transfer from the atomic sphere to the
surrounding: the larger the relaxation, the larger the charge transfer
\cite{Moreschini09b}. This mechanism determines the energy position of
the degenerate point $E_{0}$---cf.\ \eref{eq:dispersion}---and,
consequently, the Fermi energy or band-filling of the surface states
(2DEG).

A second mechanism is the atomic spin-orbit parameter. Bi and Pb are
heavy elements with large SO parameter ($\unit[1.25]{eV}$ for Bi and
$\unit[0.91]{eV}$ for Pb \cite{Wittel74}), in contrast to the lighter
element Sb ($\unit[0.4]{eV}$ \cite{Wittel74}). The Rashba splitting
depends both on the atomic SO-coupling strength and the potential
gradient \cite{Petersen00}. Since the latter should not
differ considerably among the considered systems, the spin splitting
is mainly determined by the atomic SO coupling. We expect that with
increasing Sb content, the spin splitting decreases.

A third mechanism is electron doping or band filling. Pb has one
electron less than Bi ($Z_{\mathrm{Pb}} = 82$, $Z_{\mathrm{Bi}} =
83$). Within a rigid-band model, the surface states in Pb/Ag(111) are
shifted to higher energies, as compared to those in Bi/Ag(111). This
picture is confirmed by experiments and first-principles calculations \cite{Ast08}.

\subsection{Ordered surface alloys}
The ordered surface alloys Bi/Ag(111), Pb/Ag(111), and Sb/Ag(111) have
been studied previously in detail \cite{Ast07a, Bihlmayer07, Moreschini09a}. 
They show two sets of surface states; a first set is
unoccupied and consists mainly of $p_{x}p_{y}$ orbitals (for
Bi/Cu(111), see \cite{Mirhosseini09}). In this paper, we focus on the
other set which is either completely or partially occupied and
consists of $sp_{z}$ orbitals. The effective mass $m^{\star}$ of both
sets is negative, implying a negative dispersion.

\paragraph{Sb/Ag(111).}
We address briefly the abovementioned relaxation mechanism by
considering two cases for Sb/Ag(111): (i) an Sb relaxation as
calculated by VASP ($\unit[9.6]{\%}$) and (ii) an artificial
relaxation of $\unit[25]{\%}$. The charge transfer from the Sb
muffin-tin spheres to the surrounding is increased for the larger
relaxation ($\unit[2.05]{\%}$ as compared to $\unit[0.94]{\%}$, with
respect to the nominal valence charge; cf.\ \cite{Moreschini09b}).
Consequently, the surface states are shifted towards higher energies
by $\unit[0.16]{eV}$, as obtained from the degeneracy point $E_{0}$.
Further, the spin splitting $k_{\mathrm{R}}$ becomes increased as well
($\unit[0.03]{\AA^{-1}}$ as compared to $\unit[0.02]{\AA^{-1}}$). This
corroborates that the relaxation mainly affects the crossing point
$E_{0}$ (or Fermi energy) rather than the spin splitting.

\subsection{Disordered binary alloys}
\label{sec:binary}
\begin{figure}
  \centering
  \includegraphics[width = 0.8\columnwidth]{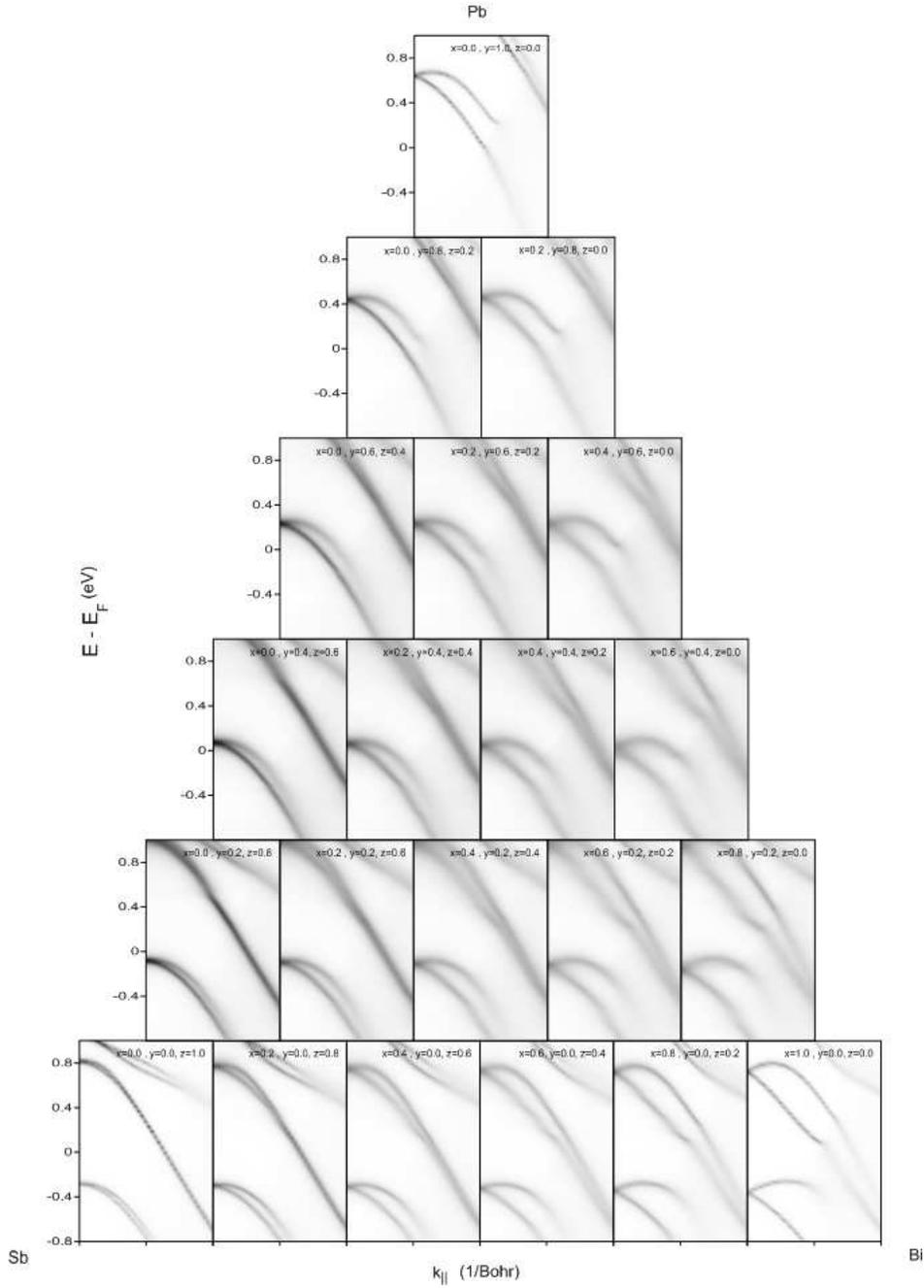}
  \caption{Surface states of disordered ternary alloys
    Bi$_{x}$Pb$_{y}$Sb$_{1-x-y}$/Ag(111) along
    $\bar{\Gamma}$--$\bar{\mathrm{K}}$ of the two-dimensional Brillouin
    zone.  The spectral density at a heavy-element site
    Bi$_{x}$Pb$_{y}$Sb$_{1-x-y}$ is depicted as linear gray scale,
    with dark gray corresponding to high spectral weight; cf.\
    \eref{eq:sd}.}
  \label{fig:rashba}
\end{figure}

\paragraph{Bi$_{x}$Pb$_{1-x}$/Ag(111).}
In the disordered binary alloy Bi$_{x}$Pb$_{1-x}$/Ag(111), which has
been studied previously \cite{Ast08}, the ratio of the Rashba energy
$E_{\mathrm{\mathrm{R}}}$ and the Fermi energy $E_{\mathrm{F}}$ can be
chosen within a wide range, in dependence on the Bi concentration $x$.
For both Bi and Pb, $\unit[0.99]{\%}$ of the atomic charge atom is
removed from the muffin-tin sphere, which is in agreement with the
close outward relaxation of Bi and Pb ($\unit[15]{\%}$ and
$\unit[18]{\%}$). As noted before, Pb has one valence electron less
than Bi, which explains the sizable shift of the surface states to
higher energies (band-filling mechanism; cf.\ the panels on the
right-hand side of \fref{fig:rashba}).  Although the relaxation is of
the same order, the splitting is smaller for Pb (topmost
panel in \fref{fig:rashba}). This can be attributed to the smaller
atomic spin-orbit parameter of Pb ($\unit[0.91]{eV}$ for Pb and $\unit[1.25]{eV}$ for Bi
\cite{Wittel74}).

\paragraph{Bi$_{x}$Sb$_{1-x}$/Ag(111).}
Recently, the surface states of the disordered binary alloys
Bi$_{x}$Sb$_{1-x}$/Ag(111) were mapped out by angle-resolved
photoelectron spectroscopy. The momentum offset $k_{\mathrm{R}}$ evolves
continuously with increasing Bi concentration $x$. The splitting
decreases sizably for $x < 0.50$ \cite{Meier.Ast.private}.

In theory, the outward relaxation of Bi is larger than for Sb
($\unit[15]{\%}$ and $\unit[9.6]{\%}$, respectively).  Consequently
the charge which is removed from the Sb sphere ($\unit[0.94]{\%}$) is
smaller than that of Bi ($\unit[0.99]{\%}$).  Since Bi and Sb are
iso-electronic, with valence-shell configuration $5p^{3}$ and
$6p^{3}$, $E_{0}$ remains almost unaffected by $x$, as can be seen in
the bottom row of \fref{fig:rashba}. The spin splitting for Sb is much
less than for Bi, in agreement with the atomic spin-orbit parameter
($\unit[0.4]{eV}$ and $\unit[1.25]{eV}$). In accord with experimental
results, the Rashba splitting $k_{\mathrm{R}}$ evolves with Bi concentration $x$.

To elucidate further the effect of the relaxation, we calculated the
splitting of Bi$_{0.6}$Sb$_{0.4}$/Ag(111) for two relaxations. The
interpolated relaxation for Bi$_{0.6}$Sb$_{0.4}$/Ag(111) is $\unit[12.8]{\%}$
(shown at $(x, y, z) = (0.6, 0.4, 0.0)$ in \fref{fig:rashba}), for the
artificial relaxed system the outward relaxation is taken as
$\unit[19]{\%}$ (not shown here). The charge transfer for the two
systems is very close, and the difference in the splitting is
negligibly small. Hence, the splitting is negligibly sensitive to the
relaxation, as was already established for Sb/Ag(111).

\paragraph{Pb$_{y}$Sb$_{1-y}$/Ag(111).}
To complete the picture of the binary alloys we turn to
Pb$_{y}$Sb$_{1-y}$/Ag(111), for which experimental results are not
available.  The trends which have been discussed before are as well
found in these alloys (cf.\ the panels on the left-hand side of
\fref{fig:rashba}).  As the Pb concentration increases, $E_{0}$ shifts
down from $E_{\mathrm{F}} + \unit[0.6]{eV}$ to $E_{\mathrm{F}} -
\unit[0.4]{eV}$, implying that the surface states become completely
filled at about $y = 0.3$. As for Bi$_{x}$Sb$_{1-x}$/Ag(111), the spin
splitting increases with $y$.

\subsection{Disordered ternary alloys Bi$_{x}$Pb$_{y}$Sb$_{1-x-y}$/Ag(111)}
\label{sec:ternary}
Having established the ingredients which are necessary for
independently tuning the Fermi energy and the spin splitting in the
surface alloys---by investigating the disordered binary surface
alloys---we now mix them to disordered ternary alloys
Bi$_{x}$Pb$_{y}$Sb$_{1-x-y}$/Ag(111). By choosing appropriate
concentrations $x$ and $y$, the degeneracy point $E_{0}$ and the
Rashba splitting are tuned. Note that the splitting
$k_{\mathrm{\mathrm{R}}}$ and the Rashba energy
$E_{\mathrm{\mathrm{R}}}$ are not fully independent; both can be
expressed (in a free-electron model) in terms of the effective
electron mass and the Rashba parameter [cf.\ \eref{eq:Deltak} and
\eref{eq:ERashba}].

In \fref{fig:rashba} the surface-state dispersions of ternary alloys
Bi$_{x}$Pb$_{y}$Sb$_{1-x-y}$/Ag(111) are shown. The concentrations $x$
and $y$ have been varied in steps of $0.2$. A common feature of the
spectral density of the binary and ternary alloys is a finite
lifetime of the spectral density, which is the consequence of the
substitutional disorder.

The Rashba characteristic of the ternary alloys follow the general
trends of the binary alloys which have been discussed before. In the
ternary alloys with larger outward relaxation (i.\,e.\ the Bi- and
Pb-rich compounds), the degenerate point $E_{0}$ shifts toward higher
energies (main mechanism: relaxation).  The larger the concentration of
heavy elements Bi and Pb as compared to the Sb concentration, the
larger the splitting $k_{\mathrm{R}}$ (main mechanism: atomic spin-orbit
parameter).  The degenerate point $E_{0}$ shifts upward with
increasing Pb concentration (main mechanism: band filling).

\begin{figure}[!b]
  \centering
  \includegraphics[width = 0.8\columnwidth]{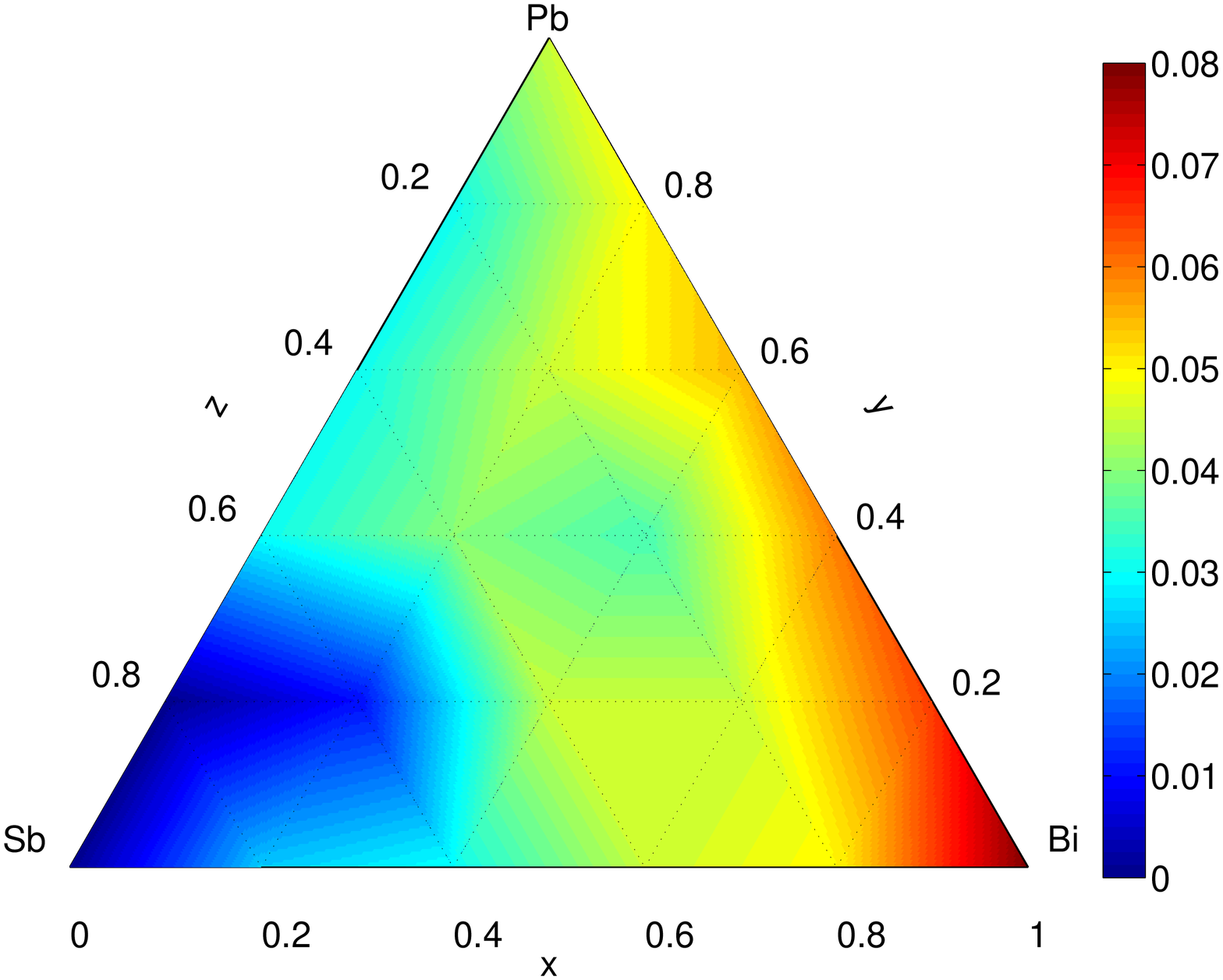}
  \includegraphics[width = 0.8\columnwidth]{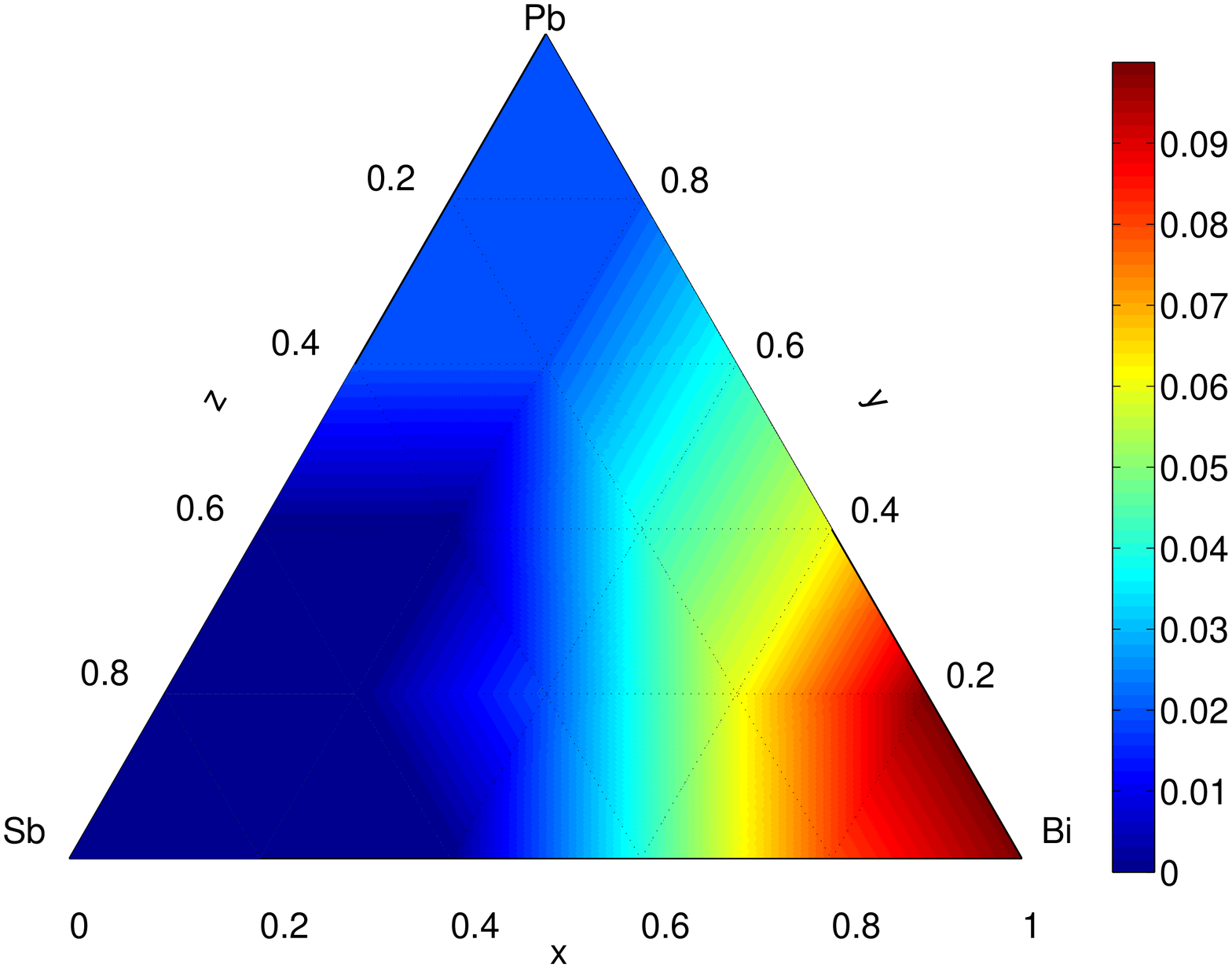}
  \caption{Spin splitting in disordered ternary alloys
    Bi$_{x}$Pb$_{y}$Sb$_{1-x-y}$/Ag(111). Top: The surface-state
    displacement $k_{\mathrm{R}}$ (in reciprocal space) is depicted as
    color scale as a function of Bi concentration $x$, Pb
    concentration $y$, and Sb concentration $z = 1 - x - y$. The color
    bar on the right is in units of $\unit{\AA^{-1}}$.  Bottom: Same
    as in the top but for the Rashba energy $E_{\mathrm{R}}$. The
    color bar is in $\unit{eV}$.}
\label{fig:figure1-3}
\end{figure}

The shift $k_{\mathrm{R}}$ of the surface states in reciprocal space
versus concentrations $x$ and $y$ is shown in \fref{fig:figure1-3}
(top). As expected, the smallest splitting (dark blue) shows up for
Sb/Ag(111) ($z = 1 - x - y = 1$), while the largest (dark red)
corresponds to Bi/Ag(111) ($x = 1$). For Pb/Ag(111), $k_{\mathrm{R}}$ is of
intermediate order (green/yellow). Surprisingly, the splitting is not
monotonic, as one might have expected in a rigid-band picture. For example, 
$k_{\mathrm{R}}$ shows a local minimum at $(x, y, z) \approx (0.4, 0.4, 0.2)$.

As $k_{\mathrm{R}}$, the Rashba energy $E_{\mathrm{R}}$ depends
monotonously in a large range of concentrations (bottom in
\fref{fig:figure1-3}). Sizable Rashba energies are found mainly for
Bi-rich alloys, say for $x > 0.5$.  This implies that for accessing
region~I, Bi-rich surface alloys are inevitable. For smaller $x$ (blue
areas in the bottom panel of \fref{fig:figure1-3}), the energy range
of region~I could be too small to be employed in experiments.

\begin{figure}[!t]
  \centering
  \includegraphics[width = 0.8\columnwidth]{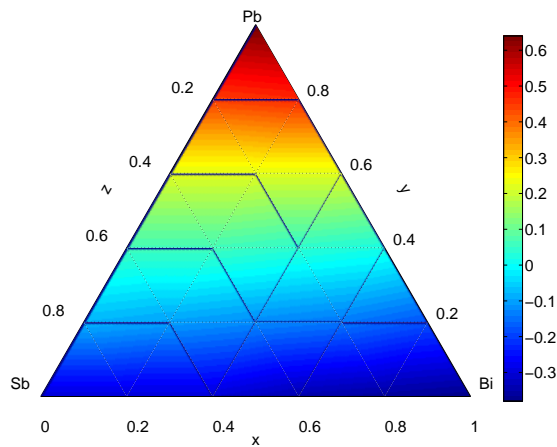}
  \caption{Surface-state energy in disordered ternary alloys
    Bi$_{x}$Pb$_{y}$Sb$_{1-x-y}$/Ag(111). The degeneracy energy
    $E_{0}$ of the surface state, with respect to the Fermi level
    $E_{\mathrm{F}}$, is depicted as color scale as a function of Bi
    concentration $x$, Pb concentration $y$, and Sb concentration $z =
    1 - x - y$. The color bar on the right is in $\unit{eV}$. At
    negative energies, the surface states are fully occupied (blue
    area).}
\label{fig:figure3}
\end{figure}

The energy $E_{0}$ of the degeneracy point depends almost linearly on
the heavy elements' concentrations $x$ and $y$ (\fref{fig:figure3}).
For equal Bi and Sb concentrations ($x = z$) it is nearly constant;
upon adding Pb, $E_{0}$ shifts up.  For systems with about
$\unit[40]{\%}$ of Pb concentration, $E_{0}$ is very close to the
Fermi level $E_{\mathrm{F}}$, so that the latter lies in region I
\cite{Moreschini09b}.

In summary, the above results support that both Fermi energy and spin splitting in
the surface states can be tuned independently, as is readily apparent
from the different shapes in \fref{fig:figure1-3} and
\fref{fig:figure3}. A very interesting region in the ternary plots is
around $(x, y, z) \approx (0.6, 0.3, 0.1)$, where the degenerate point
$E_{0}$ and the Fermi energy $E_{\mathrm{F}}$ coincide. Keeping the Sb
concentration constant and changing the Pb concentration of about
$\unit[10]{\%}$ is accompanied by transition between region I and
region II, while $k_{\mathrm{R}}$ and $E_{\mathrm{R}}$ are almost
constant. It is also possible to tune $E_{\mathrm{R}}$ and
$k_{\mathrm{R}}$ while keeping the position of degenerate point
constant. The changes of $k_{\mathrm{R}}$ and $E_{\mathrm{R}}$ are not
independent but $k_{\mathrm{R}}$ depends more sensitive on the
concentrations than $E_{\mathrm{R}}$.

\section{Conclusions}
\label{sec:conclusions}
Disordered ternary surface alloys Bi$_{x}$Pb$_{y}$Sb$_{1-x-y}$/Ag(111)
allow to fabricate a two-dimensional electron gas with specific Rashba
spin-orbit splitting and Fermi energy which can be investigated by
surface-scientific methods (scanning tunneling probes and especially
photoelectron spectroscopy). In particular, the important transition
from energy region I (that is, the Fermi energy $E_{\mathrm{F}}$ lies
above the degeneracy point $E_{0}$) to region II ($E_{\mathrm{F}}$
below $E_{0}$) can be studied for different strengths of the Rashba
spin-orbit coupling. Thus, the present study may stimulate further
experiments on Rashba systems and their unique properties.

\section{Acknowledgment}
We gratefully acknowledge very fruitful discussions with Christian Ast, Hugo J Dil, Isabella Gierz, and
Fabian Meier.

\section*{References}
\bibliographystyle{unsrt}
\bibliography{./short,./refs,./privateC}

\end{document}